\documentclass[aps,pra,showpacs,superscriptaddress,preprint]{revtex4}
\usepackage{amssymb}
\usepackage{amsmath}
\usepackage{graphicx}

\catcode`ð=\active
 \defð{\u{g}}
 \catcode`Ð=\active
 \defÐ{\u{G}}
 \catcode`Ý=\active
\defÝ{\. I}
 \catcode`ö=\active
\defö{\"{o}}
 \catcode`Ö=\active
 \defÖ{\"O}
 \catcode`ü=\active
 \defü{\"{u}}
 \catcode`Ü=\active
 \defÜ{\"{U}}
 \catcode`Þ=\active
\defÞ{\c{S}}
 \catcode`þ=\active
 \defþ{\c{s}}
 \catcode`ý=\active
 \defý{{\i}}
 \catcode`ç=\active
\defç{\d{c}}
 \catcode`Ç=\active
\defÇ{\d{C}}

\input{tcilatex}

\begin{document}

\title{Approximate solutions of the Dirac equation for the Rosen-Morse
potential including the spin-orbit centrifugal term }
\author{Sameer M. Ikhdair}
\email[E-mail: ]{sikhdair@neu.edu.tr; sikhdair@gmail.com}
\affiliation{Department of Physics, Near East University, Nicosia, North Cyprus, Turkey}
\date{%
\today%
}

\begin{abstract}
We give the approximate analytic solutions of the Dirac equations for the
Rosen-Morse potential including the spin-orbit centrifugal term. In the
framework of the spin and pseudospin symmetry concept, we obtain the
analytic bound state energy spectra and corresponding two-component upper-
and lower-spinors of the two Dirac particles, in closed form, by means of
the Nikiforov-Uvarov method. The special cases of the $s$-wave $\kappa =\pm
1 $ ($l=\widetilde{l}=0)$ Rosen-Morse potential, the Eckart-type potential,
the PT-symmetric Rosen-Morse potential and non-relativistic limits are
briefly studied.

\end{abstract}

\maketitle

\newpage

\section{Introduction}

Within the framework of the Dirac equation the spin symmetry arises if the
magnitude of the attractive scalar potential $S(r)$ and repulsive vector
potential are nearly equal, $S(r)\sim V(r)$ in the nuclei (\textit{i.e.},
when the difference potential $\Delta (r)=V(r)-S(r)=C_{s}=$ constant$).$
However, the pseudospin symmetry occurs if $S(r)\sim -V(r)$ are nearly equal
(\textit{i.e.}, when the sum potential $\Sigma (r)=V(r)+S(r)=C_{ps}=$
constant$)$ [1-3]$.$ The spin symmetry is relevant for mesons [4]. The
pseudospin symmetry concept has been applied to many systems in nuclear
physics and related areas [2-7] and used to explain features of deformed
nuclei [8], the super-deformation [9] and to establish an effective nuclear
shell-model scheme [5,6,10]. The pseudospin symmetry introduced in nuclear
theory refers to a quasi-degeneracy of the single-nucleon doublets and can
be characterized with the non-relativistic quantum numbers $(n,l,j=l+1/2)$
and $(n-1,l+2,j=l+3/2),$ where $n,$ $l$ and $j$ are the single-nucleon
radial, orbital and total angular momentum quantum numbers for a single
particle, respectively [5,6]. The total angular momentum is given as $j=%
\widetilde{l}+\widetilde{s},$ where $\widetilde{l}=l+1$ is a pseudo-angular
momentum and $\widetilde{s}=1/2$ is a pseudospin angular momentum. In real
nuclei, the pseudospin symmetry is only an approximation and the quality of
approximation depends on the pseudo-centrifugal potential and pseudospin
orbital potential [11]. Alhaidari \textit{et al}. [12] investigated in
detail the physical interpretation on the three-dimensional Dirac equation
in the context of spin symmetry limitation $\Delta (r)=0$ and pseudospin
symmetry\ limitation $\Sigma (r)=0.$

Some authors have applied the spin and pseudospin symmetry on several
physical potentials, such as the harmonic oscillator [12-15], the
Woods-Saxon potential [16], the Morse potential [17,18], the Hulth\'{e}n
potential [19], the Eckart potential [20-22], the molecular diatomic
three-parameter potential [23], the P\"{o}schl-Teller potential [24], the
Rosen-Morse potential [25] and the generalized Morse potential [26].

The exact solutions of the Dirac equation for the exponential-type
potentials are possible only for the $s$-wave ($l=0$ case). However, for $l$%
-states an approximation scheme has to be used to deal with the centrifugal
and pseudo-centrifugal terms. Many authors have used different methods to
study the partially exactly solvable and exactly solvable Schr\"{o}dinger,
Klein-Gordon (KG) and Dirac equations in $1D,3D$ and/or any $D$-dimensional
cases for different potentials [27-39]. In the context of
spatially-dependent mass, we have also used and applied a recently proposed
approximation scheme [40] for the centrifugal term to find a quasi-exact
analytic bound-state solution of the radial KG equation with
spatially-dependent effective mass for scalar and vector Hulth\'{e}n
potentials in any arbitrary dimension $D$ and orbital angular momentum
quantum number $l$ within the framework of the NU method [40-42].

Another physical potential is the Rosen-Morse potential [43] expressed in
the form%
\begin{equation}
V(r)=-V_{1}\sec h^{2}\alpha r+V_{2}\tanh \alpha r,
\end{equation}%
where $V_{1}$ and $V_{2}$ denote the depth of the potential and $\alpha $ is
the range of the potential. This potential is useful for describing
interatomic interaction of the linear molecules and helpful for disscussing
polyatomic vibration energies such as the vibration states of $NH_{3}$
molecule [43]. It is shown that the Rosen-Morse potential and its
PT-symmetric version are the special cases of the five-parameter
exponential-type potential model [44,45]. The exact energy spectrum of the
trigonometric Rosen-Morse potential has been investigated by using
supersymmetric and improved quantization rule methods [46,47].

Recently, many works have been done to solve the Dirac equation to obtain
the energy equation and the two-component spinor wave functions. Jia \textit{%
et al.} [48] employed an improved approximation scheme to deal with the
pseudo-centrifugal term to solve the Dirac equation with the generalized P%
\"{o}schl-Teller potential for arbitrary spin-orbit quantum number $\kappa .$
Zhang \textit{et al.} [49] solved the Dirac equation with equal Scarf-type
scalar and vector potentials by the method of the supersymmetric quantum
mechanics (SUSYQM), shape invariance approach and the alternative method.
Zou \textit{et al.} [50] solved the Dirac equation with equal Eckart scalar
and vector potentials in terms of SUSYQM method, shape invariance approach
and function analysis method. Wei and Dong [51] obtained approximately the
analytical bound state solutions of the Dirac equation with the
Manning-Rosen for arbitrary spin-orbit coupling quantum number $\kappa .$
Thylwe [52] presented the approach inspired by amplitude-phase method for
analyzing the radial Dirac equation to calculate phase shifts by including
the spin- and pseudo-spin symmetries of relativistic spectra. Alhaidari [53]
solved Dirac equation by separation of variables in spherical coordinates
for a large class of non-central electromagnetic potentials. Berkdemir and
Sever [54] investigated systematically the pseudospin symmetry solution of
the Dirac equation for spin $1/2$ particles moving within the Kratzer
potential connected with an angle-dependent potential. Alberto \textit{et
al. }[55] concluded that the values of energy spectra may not depend on the
spinor structure of the particle, i.e., whether one has a spin-$1/2$ or a
spin-$0$ particle. Also, they showed that a spin-$1/2$ or a spin-$0$
particle with the same mass and subject to the same scalar $S(r)$ and vector 
$V(r)$ potentials of equal magnitude, i.e., $S(r)=\pm V(r),$ will have the
same energy spectrum (isospectrality), including both bound and scattering
states.

In the present paper, our aim is to study the analytic solutions of the
Dirac equation for the Rosen-Morse potential with arbitrary spin-orbit
quantum number $\kappa $ by using a new approximation to deal with the
centrifugal term. However, we use the approximation given in Ref. [56] which
is quite different from the ones used in our previous works [39,40,42], $%
\frac{1}{r^{2}}\approx \alpha ^{2}\left[ d+\frac{e^{-\alpha r}}{\left(
1-e^{-\alpha r}\right) ^{2}}\right] $ where $d=0$ or $d=\frac{1}{12}.$ The
approximation given in [56] is convenient for the Rosen-Morse type potential
because one may propose a more reasonable physical wave functions for this
system. Under the conditions of the spin symmetry $S(r)\sim V(r)$ and
pseudospin symmetry $S(r)\sim -V(r)$, we investigate the bound state energy
eigenvalues and corresponding upper and lower spinor wave functions in the
framework of the NU method. We also show that the spin and pseudospin
symmetry Dirac solutions can be reduced to the $S(r)=V(r)$ and $S(r)=-V(r)$
in the cases of exact spin symmetry limitation $\Delta (r)=0$ and pseudospin
symmetry limitation $\Sigma (r)=0,$ respectively. Furthermore, the solutions
obtained for the Dirac equation can be easily reduced to the Schr\"{o}dinger
solutions when the appropriate map of parameters is used.

The paper is structured as follows: In Sect. 2, we outline the NU method.
Section 3 is devoted to the analytic bound state solutions of the ($3+1$%
)-dimensional Dirac equation for the Rosen-Morse quantum system obtained by
means of the NU method. The spin symmetry and pseudospin symmetry solutions
are investigated. In Sect. 4, we study the cases $\kappa =\pm 1$ ($l=%
\widetilde{l}=0,$ i.e., $s$-wave), the Eckart-type potential, the
PT-symmetric Rosen-Morse potential. Finally, the relevant conclusions are
given in Sect. 5.

\section{NU Method}

The NU method [41] is briefly outlined here. It was proposed to solve the
second-order differential equation of hypergeometric-type: 
\begin{equation}
\psi _{n}^{\prime \prime }(r)+\frac{\widetilde{\tau }(r)}{\sigma (r)}\psi
_{n}^{\prime }(r)+\frac{\widetilde{\sigma }(r)}{\sigma ^{2}(r)}\psi
_{n}(r)=0,
\end{equation}%
where $\sigma (r)$ and $\widetilde{\sigma }(r)$ are polynomials, at most, of
second-degree, and $\widetilde{\tau }(r)$ is a first-degree polynomial. In
order to find a particular solution for Eq. (2), let us decompose the
wavefunction $\psi _{n}(r)$ as follows:%
\begin{equation}
\psi _{n}(r)=\phi (r)y_{n}(r),
\end{equation}%
and use%
\begin{equation}
\left[ \sigma (r)\rho (r)\right] ^{\prime }=\tau (r)\rho (r),
\end{equation}%
to reduce Eq. (2) to the form%
\begin{equation}
\sigma (r)y_{n}^{\prime \prime }(r)+\tau (r)y_{n}^{\prime }(r)+\lambda
y_{n}(r)=0,
\end{equation}%
with%
\begin{equation}
\tau (r)=\widetilde{\tau }(r)+2\pi (r),\text{ }\tau ^{\prime }(r)<0,
\end{equation}%
where the prime denotes the differentiation with respect to $r.$ One is
looking for a family of solutions corresponding to%
\begin{equation}
\lambda =\lambda _{n}=-n\tau ^{\prime }(r)-\frac{1}{2}n\left( n-1\right)
\sigma ^{\prime \prime }(r),\ \ \ n=0,1,2,\cdots .
\end{equation}%
The $y_{n}(r)$ can be expressed in terms of the Rodrigues relation:%
\begin{equation}
y_{n}(r)=\frac{B_{n}}{\rho (r)}\frac{d^{n}}{dr^{n}}\left[ \sigma ^{n}(r)\rho
(r)\right] ,
\end{equation}%
where $B_{n}$ is the normalization constant and the weight function $\rho
(r) $ is the solution of the differential equation (4). The other part of
the wavefunction (3) must satisfy the following logarithmic equation%
\begin{equation}
\frac{\phi ^{\prime }(r)}{\phi (r)}=\frac{\pi (r)}{\sigma (r)}.
\end{equation}%
By defining 
\begin{equation}
k=\lambda -\pi ^{\prime }(r).
\end{equation}%
one obtains the polynomial

\begin{equation}
\pi (r)=\frac{1}{2}\left[ \sigma ^{\prime }(r)-\widetilde{\tau }(r)\right]
\pm \sqrt{\frac{1}{4}\left[ \sigma ^{\prime }(r)-\widetilde{\tau }(r)\right]
^{2}-\widetilde{\sigma }(r)+k\sigma (r)},
\end{equation}%
where $\pi (r)$ is a parameter at most of order $1.$ The expression under
the square root sign in the above equation can be arranged as a polynomial
of second order where its discriminant is zero. Hence, an equation for $k$
is being obtained. After solving such an equation, the $k$ values are
determined through the NU method. \ 

In this regard, we derive a parametric generalization version of the NU
method valid for any solvable potential by the method. We begin by writting
the hypergeometric equation in general parametric form as 
\begin{equation}
\left[ r\left( c_{3}-c_{4}r\right) \right] ^{2}\psi _{n}^{\prime \prime }(r)+%
\left[ r\left( c_{3}-c_{4}r\right) \left( c_{1}-c_{2}r\right) \right] \psi
_{n}^{\prime }(r)+\left( -\xi _{1}r^{2}+\xi _{2}r-\xi _{3}\right) \psi
_{n}(r)=0,
\end{equation}%
with%
\begin{equation}
\widetilde{\tau }(r)=c_{1}-c_{2}r,
\end{equation}%
\begin{equation}
\sigma (r)=r\left( c_{3}-c_{4}r\right) ,
\end{equation}%
\begin{equation}
\widetilde{\sigma }(r)=-\xi _{1}r^{2}+\xi _{2}r-\xi _{3},
\end{equation}%
where the coefficients $c_{i}$ ($i=1,2,3,4$) and the analytic expressions $%
\xi _{j}$ ($j=1,2,3$). Furthermore, in comparing Eq. (12) with the
counterpart Eq. (2), one obtains the appropriate analytic polynomials,
energy equation and wave functions together with the associated coefficients
expressed in general parameteric form as displayed in Appendix A.

\section{Analytic Solution of the Dirac-Rosen-Morse Problem}

In spherical coordinates, the Dirac equation for fermionic massive spin-$%
\frac{1}{2}$ particles interacting with arbitrary scalar potential $S(r)$
and the time-component $V(r)$ of a four-vector potential can be expressed as
[26,57-60] 
\begin{equation}
\left[ c\mathbf{\alpha }\cdot \mathbf{p+\beta }\left( Mc^{2}+S(r)\right)
+V(r)-E\right] \psi _{n\kappa }(\mathbf{r})=0,\text{ }\psi _{n\kappa }(%
\mathbf{r})=\psi (r,\theta ,\phi ),
\end{equation}%
where $E$ is the relativistic energy of the system, $M$ is the mass of a
particle, $\mathbf{p}=-i\hbar \mathbf{\nabla }$ is the momentum operator,
and $\mathbf{\alpha }$ and $\mathbf{\beta }$ are $4\times 4$ Dirac matrices,
i.e.,%
\begin{equation}
\mathbf{\alpha =}\left( 
\begin{array}{cc}
0 & \mathbf{\sigma }_{i} \\ 
\mathbf{\sigma }_{i} & 0%
\end{array}%
\right) ,\text{ }\mathbf{\beta =}\left( 
\begin{array}{cc}
\mathbf{I} & 0 \\ 
0 & -\mathbf{I}%
\end{array}%
\right) ,\text{ }\sigma _{1}\mathbf{=}\left( 
\begin{array}{cc}
0 & 1 \\ 
1 & 0%
\end{array}%
\right) ,\text{ }\sigma _{2}\mathbf{=}\left( 
\begin{array}{cc}
0 & -i \\ 
i & 0%
\end{array}%
\right) ,\text{ }\sigma _{3}\mathbf{=}\left( 
\begin{array}{cc}
1 & 0 \\ 
0 & -1%
\end{array}%
\right) ,
\end{equation}%
where $\mathbf{I}$ denotes the $2\times 2$ identity matrix and $\mathbf{%
\sigma }_{i}$ are the three-vector Pauli spin matrices. For a spherical
symmetrical nuclei, the total angular momentum operator of the nuclei $%
\mathbf{J}$ and spin-orbit matrix operator $\mathbf{K}=-\mathbf{\beta }%
\left( \mathbf{\sigma }\cdot \mathbf{L}+\mathbf{I}\right) $ commute with the
Dirac Hamiltonian, where $\mathbf{L}$ is the orbital angular momentum
operator. The spinor wavefunctions can be classified according to the radial
quantum number $n$ and the spin-orbit quantum number $\kappa $ and can be
written using the Pauli-Dirac representation in the following forms:%
\begin{equation}
\psi _{n\kappa }(\mathbf{r})=\left( 
\begin{array}{c}
f_{n\kappa }(\mathbf{r}) \\ 
g_{n\kappa }(\mathbf{r})%
\end{array}%
\right) =\frac{1}{r}\left( 
\begin{array}{c}
F_{n\kappa }(r)Y_{jm}^{l}(\theta ,\phi ) \\ 
iG_{n\kappa }(r)Y_{jm}^{\widetilde{l}}(\theta ,\phi )%
\end{array}%
\right) ,
\end{equation}%
where $F_{n\kappa }(r)$ and $G_{n\kappa }(r)$ are the radial wave functions
of the upper- and lower-spinor components, respectively and $%
Y_{jm}^{l}(\theta ,\phi )$ and $Y_{jm}^{\widetilde{l}}(\theta ,\phi )$ are
the spherical harmonic functions coupled to the total angular momentum $j$
and it's projection $m$ on the $z$ axis. The orbital and pseudo-orbital
angular momentum quantum numbers for spin symmetry $l$ and pseudospin
symmetry $\widetilde{l}$ refer to the upper- and lower-spinor components,
respectively, for which $l(l+1)=\kappa \left( \kappa +1\right) $ and $%
\widetilde{l}(\widetilde{l}+1)=\kappa \left( \kappa -1\right) $. The quantum
number $\kappa $ is related to the quantum numbers for spin symmetry $l$ and
pseudospin symmetry $\widetilde{l}$ as%
\begin{equation*}
\kappa =\left\{ 
\begin{array}{cccc}
-\left( l+1\right) =-\left( j+\frac{1}{2}\right) , & (s_{1/2},p_{3/2},\text{%
\textit{etc.}}), & \text{ }j=l+\frac{1}{2}, & \text{aligned spin }\left(
\kappa <0\right) , \\ 
+l=+\left( j+\frac{1}{2}\right) , & \text{ }(p_{1/2},d_{3/2},\text{\textit{%
etc.}}), & \text{ }j=l-\frac{1}{2}, & \text{unaligned spin }\left( \kappa
>0\right) ,%
\end{array}%
\right.
\end{equation*}%
and the quasi-degenerate doublet structure can be expressed in terms of a
pseudospin angular momentum $\widetilde{s}=1/2$ and pseudo-orbital angular
momentum $\widetilde{l}$ which is defined as 
\begin{equation*}
\kappa =\left\{ 
\begin{array}{cccc}
-\widetilde{l}=-\left( j+\frac{1}{2}\right) , & (s_{1/2},\text{ }p_{3/2},%
\text{ \textit{etc.}}), & j=\widetilde{l}-1/2, & \text{aligned spin }\left(
\kappa <0\right) , \\ 
+\left( \widetilde{l}+1\right) =+\left( j+\frac{1}{2}\right) , & \text{ }%
(d_{3/2},\text{ }f_{5/2},\text{ \textit{etc.}}), & \ j=\widetilde{l}+1/2, & 
\text{unaligned spin }\left( \kappa >0\right) ,%
\end{array}%
\right.
\end{equation*}%
where $\kappa =\pm 1,\pm 2,\cdots .$ For example, ($1s_{1/2},0d_{3/2}$) and
(2p$_{3/2},1f_{5/2}$) can be considered as pseudospin doublets.

Thus, the substitution of Eq. (18) into Eq. (16) leads to the following two
radial coupled Dirac equations for the spinor components 
\begin{subequations}
\begin{equation}
\left( \frac{d}{dr}+\frac{\kappa }{r}\right) F_{n\kappa }(r)=\left(
Mc^{2}+E_{n\kappa }-\Delta (r)\right) G_{n\kappa }(r),
\end{equation}%
\begin{equation}
\left( \frac{d}{dr}-\frac{\kappa }{r}\right) G_{n\kappa }(r)=\left(
Mc^{2}-E_{n\kappa }+\Sigma (r)\right) F_{n\kappa }(r),
\end{equation}%
where $\Delta (r)=V(r)-S(r)$ and $\Sigma (r)=V(r)+S(r)$ are the difference
and sum potentials, respectively.

Under the spin symmetry ( i.e., $\Delta (r)=C_{s}=$ constant), one can
eliminate $G_{n\kappa }(r)$ in Eq. (19a), with the aid of Eq. (19b), to
obtain a second-order differential equation for the upper-spinor component
as follows [16,26]: 
\end{subequations}
\begin{equation*}
\left[ -\frac{d^{2}}{dr^{2}}+\frac{\kappa \left( \kappa +1\right) }{r^{2}}+%
\frac{1}{\hbar ^{2}c^{2}}\left( Mc^{2}+E_{n\kappa }-C_{s}\right) \Sigma (r)%
\right] F_{n\kappa }(r)
\end{equation*}%
\begin{equation}
=\frac{1}{\hbar ^{2}c^{2}}\left( E_{n\kappa }^{2}-M^{2}c^{4}+C_{s}\left(
Mc^{2}-E_{n\kappa }\right) \right) F_{n\kappa }(r),
\end{equation}%
where $\kappa \left( \kappa +1\right) =l\left( l+1\right) ,$ $\kappa =l$ for 
$\kappa <0$ and $\kappa =-\left( l+1\right) $ for $\kappa >0.$ The spin
symmetry energy eigenvalues depend on $n$ and $\kappa ,$ \textit{i.e.}, $%
E_{n\kappa }=E(n,\kappa \left( \kappa +1\right) ).$ For $l\neq 0,$ the
states with $j=l\pm 1/2$ are degenerate. Further, the lower-spinor component
can be obtained from Eq. (19a) as%
\begin{equation}
G_{n\kappa }(r)=\frac{1}{Mc^{2}+E_{n\kappa }-C_{s}}\left( \frac{d}{dr}+\frac{%
\kappa }{r}\right) F_{n\kappa }(r),
\end{equation}%
where $E_{n\kappa }\neq -Mc^{2},$ only real positive energy states exist
when $C_{s}=0$ (exact spin symmetry).

On the other hand, under the pseudospin symmetry ( i.e., $\Sigma (r)=C_{ps}=$
constant), one can eliminate $F_{n\kappa }(r)$ in Eq. (19b), with the aid of
Eq. (19a), to obtain a second-order differential equation for the
lower-spinor component as follows [16,26]:%
\begin{equation*}
\left[ -\frac{d^{2}}{dr^{2}}+\frac{\kappa \left( \kappa -1\right) }{r^{2}}-%
\frac{1}{\hbar ^{2}c^{2}}\left( Mc^{2}-E_{n\kappa }+C_{ps}\right) \Delta (r)%
\right] G_{n\kappa }(r)
\end{equation*}%
\begin{equation}
=\frac{1}{\hbar ^{2}c^{2}}\left( E_{n\kappa }^{2}-M^{2}c^{4}-C_{ps}\left(
Mc^{2}+E_{n\kappa }\right) \right) G_{n\kappa }(r),
\end{equation}%
and the upper-spinor component $F_{n\kappa }(r)$ is obtained from Eq. (19b)
as%
\begin{equation}
F_{n\kappa }(r)=\frac{1}{Mc^{2}-E_{n\kappa }+C_{ps}}\left( \frac{d}{dr}-%
\frac{\kappa }{r}\right) G_{n\kappa }(r),
\end{equation}%
where $E_{n\kappa }\neq Mc^{2},$ only real negative energy states exist when 
$C_{ps}=0$ (exact pseudospin symmetry). From the above equations, the energy
eigenvalues depend on the quantum numbers $n$ and $\kappa $, and also the
pseudo-orbital angular quantum number $\widetilde{l}$ according to $\kappa
(\kappa -1)=\widetilde{l}(\widetilde{l}+1),$ which implies that $j=%
\widetilde{l}\pm 1/2$ are degenerate for $\widetilde{l}\neq 0.$ The quantum
condition is obtained from the finiteness of the solution at infinity and at
the origin point, i.e., $F_{n\kappa }(0)=G_{n\kappa }(0)=0$ and $F_{n\kappa
}(\infty )=G_{n\kappa }(\infty )=0.$

At this stage, we take the vector and scalar potentials in the form of
Rosen-Morse potential model (see Eq. (1)). Equations (20) and (22) can be
solved exactly for $\kappa =0,-1$ and $\kappa =0,1,$ respectively, because
of the spin-orbit centrifugal and pseudo-centrifugal terms. Therefore, to
find approximate solution for the radial Dirac equation with the Rosen-Morse
potential, we have to use an approximation for the spin-orbit centrifugal
term. For values of $\kappa $ that are not large and vibrations of the small
amplitude about the minimum, Lu [56] has introduced an approximation to the
centrifugal term near the minimum point $r=r_{e}$ as 
\begin{equation}
\frac{1}{r^{2}}\approx \frac{1}{r_{e}^{2}}\left[ D_{0}+D_{1}\frac{-\exp
(-2\alpha r)}{1+\exp (-2\alpha r)}+D_{2}\left( \frac{-\exp (-2\alpha r)}{%
1+\exp (-2\alpha r)}\right) ^{2}\right] ,
\end{equation}%
where 
\begin{equation*}
D_{0}=1-\left( \frac{1+\exp (-2\alpha r_{e})}{2\alpha r_{e}}\right)
^{2}\left( \frac{8\alpha r_{e}}{1+\exp (-2\alpha r_{e})}-\left( 3+2\alpha
r_{e}\right) \right) ,
\end{equation*}%
\begin{equation*}
D_{1}=-2\left( \exp (2\alpha r_{e})+1\right) \left[ 3\left( \frac{1+\exp
(-2\alpha r_{e})}{2\alpha r_{e}}\right) -\left( 3+2\alpha r_{e}\right)
\left( \frac{1+\exp (-2\alpha r_{e})}{2\alpha r_{e}}\right) \right] ,
\end{equation*}%
\begin{equation}
D_{2}=\left( \exp (2\alpha r_{e})+1\right) ^{2}\left( \frac{1+\exp (-2\alpha
r_{e})}{2\alpha r_{e}}\right) ^{2}\left( 3+2\alpha r_{e}-\frac{4\alpha r_{e}%
}{1+\exp (-2\alpha r_{e})}\right) ,
\end{equation}%
and higher order terms are neglected.

\subsection{Spin symmetry solution of the Rosen-Morse Problem}

We take the sum potential in Eq. (20) as the Rosen-Morse potential model,
i.e.,%
\begin{equation}
\Sigma (r)=-4V_{1}\frac{\exp (-2\alpha r)}{\left( 1+\exp (-2\alpha r)\right)
^{2}}+V_{2}\frac{\left( 1-\exp (-2\alpha r)\right) }{\left( 1+\exp (-2\alpha
r)\right) }.
\end{equation}%
The choice of $\Sigma (r)=2V(r)\rightarrow V(r)$ as mentioned in Ref. [12]
enables one to reduce the resulting relativistic solutions into their
non-relativistic limit under appropriate transformations. $.$

Using the approximation given by Eq. (24) and introducing a new parameter
change $z(r)=-\exp (-2\alpha r)$, this allows us to decompose the
spin-symmetric Dirac equation (20) into the Schr\"{o}dinger-like equation in
the spherical coordinates for the upper-spinor component $F_{n\kappa }(r),$%
\begin{equation}
\left[ \frac{d^{2}}{dz^{2}}+\frac{\left( 1-z\right) }{z\left( 1-z\right) }%
\frac{d}{dz}+\frac{\left( -\beta _{1}z^{2}+\beta _{2}z-\varepsilon _{n\kappa
}^{2}\right) }{z^{2}\left( 1-z\right) ^{2}}\right] F_{n\kappa }(z)=0,\text{ }%
F_{n\kappa }(0)=F_{n\kappa }(1)=0,
\end{equation}%
with 
\begin{subequations}
\begin{equation}
\text{ }\varepsilon _{n\kappa }=\frac{1}{2\alpha }\sqrt{\frac{\omega }{%
r_{e}^{2}}D_{0}+\frac{1}{\hbar ^{2}c^{2}}\left( Mc^{2}+E_{n\kappa
}-C_{s}\right) \left( Mc^{2}-E_{n\kappa }+V_{2}\right) }>0,
\end{equation}%
\begin{equation}
\beta _{1}=\frac{1}{4\alpha ^{2}}\left\{ \frac{\omega }{r_{e}^{2}}\left(
D_{0}+D_{1}+D_{2}\right) +\frac{1}{\hbar ^{2}c^{2}}\left( Mc^{2}+E_{n\kappa
}-C_{s}\right) \left( Mc^{2}-E_{n\kappa }-V_{2}\right) \right\} ,\text{ }
\end{equation}%
\begin{equation}
\beta _{2}=\frac{1}{4\alpha ^{2}}\left\{ \frac{\omega }{r_{e}^{2}}\left(
2D_{0}+D_{1}\right) +\frac{2}{\hbar ^{2}c^{2}}\left( Mc^{2}+E_{n\kappa
}-C_{s}\right) \left( Mc^{2}-E_{n\kappa }-2V_{1}\right) \right\} ,\text{ }
\end{equation}%
where $\omega =\kappa \left( \kappa +1\right) .$

In order to solve Eq. (27) by means of the NU method, we should compare it
with Eq. (2) to obtain the following particular values for the parameters: 
\end{subequations}
\begin{equation}
\widetilde{\tau }(z)=1-z,\text{\ }\sigma (z)=z\left( 1-z\right) ,\text{\ }%
\widetilde{\sigma }(z)=-\beta _{1}z^{2}+\beta _{2}z-\varepsilon _{n\kappa
}^{2}.
\end{equation}%
Comparing Eqs. (13)-(15) with Eq. (29), we can easily obtain the
coefficients $c_{i}$ ($i=1,2,3,4$) and the analytic expressions $\xi _{j}$ ($%
j=1,2,3$). However, the values of the coefficients $c_{i}$ ($i=5,6,\cdots
,16 $) are found from the relations A1-A5 of Appendix A. Therefore, the
specific values of the coefficients $c_{i}$ ($i=1,2,\cdots ,16$) together
with $\xi _{j}$ ($j=1,2,3$) are displayed in Table 1. From the relations A6
and A7 of Appendix A together with the coefficients in Table 1, the selected
forms of $\pi (z)$ and $k$ take the following particular values 
\begin{equation}
\pi (z)=\varepsilon _{n\kappa }-\left( 1+\varepsilon _{n\kappa }+\delta
\right) z,
\end{equation}%
\begin{equation}
k=\beta _{2}-\left[ 2\varepsilon _{n\kappa }^{2}+\left( 2\delta +1\right)
\varepsilon _{n\kappa }\right] ,
\end{equation}%
respectively, where%
\begin{equation}
\delta =\frac{1}{2}\left( -1+\sqrt{1+\frac{\omega D_{2}}{\alpha ^{2}r_{e}^{2}%
}+\frac{4V_{1}}{\alpha ^{2}\hbar ^{2}c^{2}}\left( Mc^{2}+E_{n\kappa
}-C_{s}\right) }\right) ,
\end{equation}%
for bound state solutions. According to the NU method, the relations A8 and
A9 of Appendix A give%
\begin{equation*}
\tau (z)=1+2\varepsilon _{n\kappa }-\left( 3+2\varepsilon _{n\kappa
}+2\delta \right) z,
\end{equation*}%
\begin{equation}
\text{ }\tau ^{\prime }(r)=-\left( 3+2\varepsilon _{n\kappa }+2\delta
\right) <0,
\end{equation}%
with prime denotes the derivative with respect to $z.$ In addition, the
relation A10 of Appendix A gives the energy equation for the Rosen-Morse
potential in the Dirac theory as%
\begin{equation*}
\left( Mc^{2}+E_{n\kappa }-C_{s}\right) \left( Mc^{2}-E_{n\kappa
}+V_{2}\right) =-\frac{\omega D_{0}}{r_{e}^{2}}\hbar ^{2}c^{2}
\end{equation*}%
\begin{equation}
+\alpha ^{2}\hbar ^{2}c^{2}\left[ \frac{-\frac{V_{2}}{2\alpha ^{2}\hbar
^{2}c^{2}}\left( Mc^{2}+E_{n\kappa }-C_{s}\right) +\frac{\omega \left(
D_{1}+D_{2}\right) }{4\alpha ^{2}r_{e}^{2}}}{\left( n+\delta +1\right) }%
-\left( n+\delta +1\right) \right] ^{2}.
\end{equation}%
Further, for the exact spin symmetry case, $V(r)=S(r)$ or $C_{s}=0$, we
obtain%
\begin{equation*}
\left( Mc^{2}+E_{n\kappa }\right) \left( Mc^{2}-E_{n\kappa }+V_{2}\right) =-%
\frac{\omega D_{0}}{r_{e}^{2}}\hbar ^{2}c^{2}
\end{equation*}%
\begin{equation}
+\alpha ^{2}\hbar ^{2}c^{2}\left[ \frac{-\frac{V_{2}}{2\alpha ^{2}\hbar
^{2}c^{2}}\left( Mc^{2}+E_{n\kappa }-C_{s}\right) +\frac{\omega \left(
D_{1}+D_{2}\right) }{4\alpha ^{2}r_{e}^{2}}}{\left( n+\widetilde{\delta }%
+1\right) }-\left( n+\widetilde{\delta }+1\right) \right] ^{2},
\end{equation}%
with%
\begin{equation}
\widetilde{\delta }=\delta (C_{s}\rightarrow 0).
\end{equation}%
Let us now find the corresponding wave functions for this model. Referring
to Table 1 and the relations A11 and A12 of Appendix A, we find the
functions:%
\begin{equation}
\rho (z)=z^{2\varepsilon _{n\kappa }}\left( 1-z\right) ^{2\delta +1},
\end{equation}%
\begin{equation}
\phi (r)=z^{\varepsilon _{n\kappa }}\left( 1-z\right) ^{\delta +1}.
\end{equation}%
Hence, the relation A13 of Appendix A gives%
\begin{equation}
y_{n}(z)=A_{n}z^{-2\varepsilon _{n\kappa }}\left( 1-z\right) ^{-\left(
2\delta +1\right) }\frac{d^{n}}{dz^{n}}\left[ z^{n+2\varepsilon _{n\kappa
}}\left( 1-z\right) ^{n+2\delta +1}\right] \sim P_{n}^{\left( 2\varepsilon
_{n\kappa },2\delta +1\right) }(1-2z),\text{ }z\in \lbrack 0,1],
\end{equation}%
where the Jacobi polynomial $P_{n}^{\left( \mu ,\nu \right) }(x)$ is defined
only for $\mu >-1,$ $\nu >-1,$ and for the argument $x\in \left[ -1,+1\right]
.$ By using $F_{n\kappa }(z)=\phi (z)y_{n}(z),$ we get the radial
upper-spinor wave functions from the relation A14 as%
\begin{equation*}
F_{n\kappa }(z)=\mathcal{N}_{n\kappa }z^{\varepsilon _{n\kappa }}\left(
1-z\right) ^{\delta +1}P_{n}^{\left( 2\varepsilon _{n\kappa },2\delta
+1\right) }(1-2z)
\end{equation*}%
\begin{equation*}
=\mathcal{N}_{n\kappa }\left( \exp (-2\alpha r)\right) ^{\varepsilon
_{n\kappa }}\left( 1-\exp (-2\alpha r)\right) ^{\delta +1}
\end{equation*}%
\begin{equation}
\times 
\begin{array}{c}
_{2}F_{1}%
\end{array}%
\left( -n,n+2\left( \varepsilon _{n\kappa }+\delta +1\right) ;2\varepsilon
_{n\kappa }+1;-\exp (-2\alpha r)\right) .
\end{equation}%
The above upper-spinor component satisfies the restriction condition for the
bound states, \textit{i.e.}, $\delta >0$ and $\varepsilon _{n\kappa }>0.$
The normalization constants $\mathcal{N}_{n\kappa }$ are calculated in
Appendix B.

Before presenting the corresponding lower-component $G_{n\kappa }(r),$ let
us recall a recurrence relation of hypergeometric function, which is used to
solve Eq. (21) and present the corresponding lower component $G_{n\kappa
}(r),$%
\begin{equation}
\frac{d}{dz}\left[ 
\begin{array}{c}
_{2}F_{1}%
\end{array}%
\left( a;b;c;z\right) \right] =\left( \frac{ab}{c}\right) 
\begin{array}{c}
_{2}F_{1}%
\end{array}%
\left( a+1;b+1;c+1;z\right) ,
\end{equation}%
with which the corresponding lower component $G_{n\kappa }(r)$ can be
obtained as follows%
\begin{equation*}
G_{n\kappa }(r)=\frac{\mathcal{N}_{n\kappa }\left( -\exp (-2\alpha r)\right)
^{\varepsilon _{n\kappa }}(1+\exp (-2\alpha r))^{\delta +1}}{\left(
Mc^{2}+E_{n\kappa }-C_{s}\right) }\left[ -2\alpha \varepsilon _{n\kappa }-%
\frac{2\alpha \left( \delta +1\right) \exp (-2\alpha r)}{\left( 1+\exp
(-2\alpha r)\right) }+\frac{\kappa }{r}\right]
\end{equation*}%
\begin{equation*}
\times 
\begin{array}{c}
_{2}F_{1}%
\end{array}%
\left( -n,n+2\left( \varepsilon _{n\kappa }+\delta +1\right) ;2\varepsilon
_{n\kappa }+1;-\exp (-2\alpha r)\right)
\end{equation*}%
\begin{equation*}
+\mathcal{N}_{n\kappa }\left[ \frac{2\alpha n\left[ n+2\left( \varepsilon
_{n\kappa }+\delta +1\right) \right] \left( -\exp (-2\alpha r)\right)
^{\varepsilon _{n\kappa }+1}\left( 1+\exp (-2\alpha r)\right) ^{\delta +1}}{%
\left( 2\varepsilon _{n\kappa }+1\right) \left( Mc^{2}+E_{n\kappa
}-C_{s}\right) }\right]
\end{equation*}%
\begin{equation}
\times 
\begin{array}{c}
_{2}F_{1}%
\end{array}%
\left( -n+1;n+2\left( \varepsilon _{n\kappa }+\delta +\frac{3}{2}\right)
;2\left( \varepsilon _{n\kappa }+1\right) ;-\exp (-2\alpha r)\right) ,
\end{equation}%
where $E_{n\kappa }\neq -Mc^{2}$ for exact spin symmetry. Here, it should be
noted that the hypergeometric series $%
\begin{array}{c}
_{2}F_{1}%
\end{array}%
\left( -n,n+2\left( \varepsilon _{n\kappa }+\delta +1\right) ;2\varepsilon
_{n\kappa }+1;-\exp (-2\alpha r)\right) $ does not terminate for $n=0$ and
thus does not diverge for all values of real parameters $\delta $ and $%
\varepsilon _{n\kappa }.$

For $C_{s}>Mc^{2}+E_{n\kappa }$ and $E_{n\kappa }<Mc^{2}+V_{2}$ or $%
C_{s}<Mc^{2}+E_{n\kappa }$ and $E_{n\kappa }>Mc^{2}+V_{2},$ we note that
parameters given in Eq. (28a) turn to be imaginary, i.e., $\varepsilon
_{n\kappa }^{2}<0$ in the $s$-state ($\kappa =-1$)$.$ As a result, the
condition of existing bound states are $\varepsilon _{n\kappa }>0$ and $%
\delta >0,$ that is to say, in the case of $C_{s}<Mc^{2}+E_{n\kappa }$ and $%
E_{n\kappa }<Mc^{2}+V_{2},$ bound-states do exist for some quantum number $%
\kappa $ such as the $s$-state ($\kappa =-1$)$.$ Of course, if these
conditions are satisfied for existing bound-states, the energy equation and
wave functions are the same as these given in Eq. (34) and Eqs. (40) and
(42).

\subsection{Pseudospin symmetry solution of the Rosen-Morse Problem}

Now taking the difference potential in Eq. (22) as the Rosen-Morse potential
model, i.e.,%
\begin{equation}
\Delta (r)=-4V_{1}\frac{\exp (-2\alpha r)}{\left( 1+\exp (-2\alpha r)\right)
^{2}}+V_{2}\frac{\left( 1-\exp (-2\alpha r)\right) }{\left( 1+\exp (-2\alpha
r)\right) },
\end{equation}%
leads us to obtain a Schr\"{o}dinger-like equation for the lower-spinor
component $G_{n\kappa }(r),$%
\begin{equation}
\left[ \frac{d^{2}}{dz^{2}}+\frac{\left( 1-z\right) }{z\left( 1-z\right) }%
\frac{d}{dz}+\frac{\left( -\widetilde{\beta }_{1}z^{2}+\widetilde{\beta }%
_{2}z-\widetilde{\varepsilon }_{n\kappa }^{2}\right) }{z^{2}\left(
1-z\right) ^{2}}\right] G_{n\kappa }(z)=0,\text{ }G_{n\kappa }(0)=G_{n\kappa
}(1)=0,
\end{equation}%
where 
\begin{subequations}
\begin{equation}
\text{ }\widetilde{\varepsilon }_{n\kappa }=\frac{1}{2\alpha }\sqrt{\frac{%
\widetilde{\omega }}{r_{e}^{2}}D_{0}-\frac{1}{\hbar ^{2}c^{2}}\left[
E_{n\kappa }^{2}-M^{2}c^{4}-\left( Mc^{2}+E_{n\kappa }\right) C_{ps}+\left(
Mc^{2}-E_{n\kappa }+C_{ps}\right) V_{2}\right] }>0,
\end{equation}%
\begin{equation}
\widetilde{\beta }_{1}=\frac{1}{4\alpha ^{2}}\left\{ \frac{\widetilde{\omega 
}}{r_{e}^{2}}\left( D_{0}+D_{1}+D_{2}\right) -\frac{1}{\hbar ^{2}c^{2}}\left[
E_{n\kappa }^{2}-M^{2}c^{4}-\left( Mc^{2}+E_{n\kappa }\right) C_{ps}-\left(
Mc^{2}-E_{n\kappa }+C_{ps}\right) V_{2}\right] \right\} ,\text{ }
\end{equation}%
\begin{equation}
\widetilde{\beta }_{2}=\frac{1}{4\alpha ^{2}}\left\{ \frac{\widetilde{\omega 
}}{r_{e}^{2}}\left( 2D_{0}+D_{1}\right) -\frac{2}{\hbar ^{2}c^{2}}\left[
E_{n\kappa }^{2}-M^{2}c^{4}-\left( Mc^{2}+E_{n\kappa }\right) C_{ps}-2\left(
Mc^{2}-E_{n\kappa }+C_{ps}\right) V_{1}\right] \right\} ,\text{ }
\end{equation}%
and $\widetilde{\omega }=\kappa \left( \kappa -1\right) .$ To avoid
repetition in the solution of Eq. (44), a first inspection for the
relationship between the present set of parameters $(\widetilde{\varepsilon }%
_{n\kappa },\widetilde{\beta }_{1},\widetilde{\beta }_{2})$ and the previous
set $(\varepsilon _{n\kappa },\beta _{1},\beta _{2})$ tells us that the
negative energy solution for pseudospin symmetry, where $S(r)=-V(r),$ can be
obtained directly from those of the positive energy solution above for spin
symmetry using the parameter map [57-59]: 
\end{subequations}
\begin{equation}
F_{n\kappa }(r)\leftrightarrow G_{n\kappa }(r),V(r)\rightarrow -V(r)\text{
(or }V_{1}\rightarrow -V_{1}\text{ and }V_{2}\rightarrow -V_{2}\text{)},%
\text{ }E_{n\kappa }\rightarrow -E_{n\kappa }\text{ and }C_{s}\rightarrow
-C_{ps}.
\end{equation}%
Following the previous results with the above transformations, we finally
arrive at the energy equation

\begin{equation*}
\left( Mc^{2}-E_{n\kappa }+C_{ps}\right) \left( Mc^{2}+E_{n\kappa
}-V_{2}\right) =-\frac{\widetilde{\omega }D_{0}}{r_{e}^{2}}\hbar ^{2}c^{2}
\end{equation*}%
\begin{equation}
+\alpha ^{2}\hbar ^{2}c^{2}\left[ \frac{\frac{V_{2}}{2\alpha ^{2}\hbar
^{2}c^{2}}\left( Mc^{2}-E_{n\kappa }+C_{ps}\right) +\frac{\widetilde{\omega }%
\left( D_{1}+D_{2}\right) }{4\alpha ^{2}r_{e}^{2}}}{\left( n+\delta
_{1}+1\right) }-\left( n+\delta _{1}+1\right) \right] ^{2},
\end{equation}%
where%
\begin{equation}
\delta _{1}=\frac{1}{2}\left( -1+\sqrt{1+\frac{\widetilde{\omega }D_{2}}{%
\alpha ^{2}r_{e}^{2}}-\frac{4V_{1}}{\alpha ^{2}\hbar ^{2}c^{2}}\left(
Mc^{2}-E_{n\kappa }+C_{ps}\right) }\right) .
\end{equation}%
By using $G_{n\kappa }(z)=\phi (z)y_{n}(z),$ we get the radial lower-spinor
wave functions as%
\begin{equation}
G_{n\kappa }(r)=\widetilde{\mathcal{N}}_{n\kappa }\left( \exp (-2\alpha
r)\right) ^{\widetilde{\varepsilon }_{n\kappa }}\left( 1-\exp (-2\alpha
r)\right) ^{\delta _{1}+1}P_{n}^{\left( 2\widetilde{\varepsilon }_{n\kappa
},2\delta _{1}+1\right) }(1-2\exp (-2\alpha r)).
\end{equation}%
The above upper-spinor component satisfies the restriction condition for the
bound states, \textit{i.e.}, $\delta _{1}>0$ and $\widetilde{\varepsilon }%
_{n\kappa }>0.$ The normalization constants $\widetilde{\mathcal{N}}_{nl}$
are calculated in Appendix B.

\section{Discussions}

In this section, we are going to study four special cases of the energy
eigenvalues given by Eqs. (34) and (47) for the spin and pseudospin
symmetry, respectively. First, let us study $s$-wave case $l=0$ ($\kappa =-1$%
) and $\widetilde{l}=0$ ($\kappa =1$) case%
\begin{equation}
\left( Mc^{2}+E_{n,-1}-C_{s}\right) \left( Mc^{2}-E_{n,-1}+V_{2}\right)
=\alpha ^{2}\hbar ^{2}c^{2}\left[ \frac{\frac{V_{2}}{2\alpha ^{2}\hbar
^{2}c^{2}}\left( Mc^{2}+E_{n,-1}-C_{s}\right) }{n+\delta _{2}+1}+n+\delta
_{2}+1\right] ^{2},
\end{equation}%
where%
\begin{equation}
\delta _{2}=\frac{1}{2}\left( -1+\sqrt{1+\frac{4V_{1}}{\alpha ^{2}\hbar
^{2}c^{2}}\left( Mc^{2}+E_{n,-1}-C_{s}\right) }\right) .
\end{equation}%
If one sets $C_{s}=0$ into Eq. (50) and $C_{ps}=0$ into Eq. (47), we obtain
for spin and pseudospin symmetric Dirac theory,%
\begin{equation}
\left( Mc^{2}+E_{n,-1}\right) \left( Mc^{2}-E_{n,-1}+V_{2}\right) =\alpha
^{2}\hbar ^{2}c^{2}\left[ \frac{\frac{V_{2}}{2\alpha ^{2}\hbar ^{2}c^{2}}%
\left( Mc^{2}+E_{n,-1}\right) }{n+\delta _{-1}+1}+n+\delta _{-1}+1\right]
^{2},
\end{equation}%
\begin{equation}
\delta _{-1}=\frac{1}{2}\left( -1+\sqrt{1+\frac{4V_{1}}{\alpha ^{2}\hbar
^{2}c^{2}}\left( Mc^{2}+E_{n,-1}\right) }\right) ,
\end{equation}%
and%
\begin{equation}
\left( Mc^{2}-E_{n,+1}\right) \left( Mc^{2}+E_{n,+1}-V_{2}\right) =+\alpha
^{2}\hbar ^{2}c^{2}\left[ -\frac{\frac{V_{2}}{2\alpha ^{2}\hbar ^{2}c^{2}}%
\left( Mc^{2}-E_{n,+1}\right) }{n+\delta _{+1}+1}+n+\delta _{+1}+1\right]
^{2},
\end{equation}%
\begin{equation}
\delta _{+1}=\frac{1}{2}\left( -1+\sqrt{1-\frac{4V_{1}}{\alpha ^{2}\hbar
^{2}c^{2}}\left( Mc^{2}-E_{n,+1}\right) }\right) .
\end{equation}%
respectively. The above solutions for the $s$-wave are found to be identical
for spin and pseudospin cases $S(r)=V(r)$ and $S(r)=-V(r),$ respectively.

Second, when we set $V_{1}\rightarrow -V_{1}$ and $V_{2}\rightarrow -V_{2}$
the potential reduces to the Eckart-type potential and energy eigenvalues
are given by%
\begin{equation}
\left( Mc^{2}+E_{n,-1}\right) \left( Mc^{2}-E_{n,-1}-V_{2}\right) =\alpha
^{2}\hbar ^{2}c^{2}\left[ -\frac{\frac{V_{2}}{2\alpha ^{2}\hbar ^{2}c^{2}}%
\left( Mc^{2}+E_{n,-1}\right) }{n+\delta _{-1}+1}+n+\delta _{-1}+1\right]
^{2},
\end{equation}%
\begin{equation}
\delta _{-1}=\frac{1}{2}\left( -1+\sqrt{1-\frac{4V_{1}}{\alpha ^{2}\hbar
^{2}c^{2}}\left( Mc^{2}+E_{n,-1}\right) }\right) ,
\end{equation}%
for spin symmetry and%
\begin{equation}
\left( Mc^{2}-E_{n,+1}\right) \left( Mc^{2}+E_{n,+1}+V_{2}\right) =+\alpha
^{2}\hbar ^{2}c^{2}\left[ \frac{\frac{V_{2}}{2\alpha ^{2}\hbar ^{2}c^{2}}%
\left( Mc^{2}-E_{n,+1}\right) }{n+\delta _{+1}+1}+n+\delta _{+1}+1\right]
^{2},
\end{equation}%
\begin{equation}
\delta _{+1}=\frac{1}{2}\left( -1+\sqrt{1+\frac{4V_{1}}{\alpha ^{2}\hbar
^{2}c^{2}}\left( Mc^{2}-E_{n,+1}\right) }\right) ,
\end{equation}%
for pseudospin symmetry.

Third, let us now discuss the non-relativistic limit of the energy
eigenvalues and wave functions of our solution. If we take $C_{s}=0$ and put 
$S(r)=V(r)=\Sigma (r),$ the non-relativistic limit of energy equation (34)
and wave functions (40) under the following appropriate transformations $%
\left( Mc^{2}+E_{n\kappa }\right) /\hbar ^{2}c^{2}\rightarrow 2\mu /\hbar
^{2}$ and $Mc^{2}-E_{n\kappa }\rightarrow -E_{nl}$ [26,57,42] become%
\begin{equation}
E_{nl}=V_{2}+\frac{\omega D_{0}}{2\mu r_{e}^{2}}\hbar ^{2}-\frac{\hbar ^{2}}{%
2\mu }\alpha ^{2}\left[ \frac{\frac{\mu }{\alpha ^{2}\hbar ^{2}}\left(
2V_{1}+V_{2}\right) -\frac{\omega D_{1}}{4\alpha ^{2}r_{e}^{2}}+\left(
n+1\right) ^{2}+\left( 2n+1\right) \widetilde{\delta }_{0}}{n+\widetilde{%
\delta }_{0}+1}\right] ^{2},
\end{equation}%
with%
\begin{equation}
\widetilde{\delta }_{0}=\frac{1}{2}\left( -1+\sqrt{1+\frac{8\mu V_{1}}{%
\alpha ^{2}\hbar ^{2}}+\frac{\omega D_{2}}{\alpha ^{2}r_{e}^{2}}}\right) ,
\end{equation}%
and the associated wave functions are%
\begin{equation}
F_{nl}(r)=\mathcal{N}_{nl}\left( \exp (-2\alpha r)\right) ^{\varepsilon
_{n\kappa }}\left( 1-\exp (-2\alpha r)\right) ^{1+\widetilde{\delta }%
_{0}}P_{n}^{\left( 2\varepsilon _{n\kappa },2\widetilde{\delta }%
_{0}+1\right) }(1-2\exp (-2\alpha r)),
\end{equation}%
where 
\begin{subequations}
\begin{equation}
\text{ }\varepsilon _{nl}=\frac{1}{2\alpha }\sqrt{\frac{\omega }{r_{e}^{2}}%
D_{0}+\frac{2\mu }{\hbar ^{2}}\left( V_{2}-E_{nl}\right) }>0,\text{ }\omega
=l(l+1),
\end{equation}%
which are identical with Ref. [25] in the solution of the Schr\"{o}dinger
equation. Finally, the Jacobi polynomials can be expressed in terms of the
hypergeometric function as 
\end{subequations}
\begin{equation}
P_{n}^{\left( \mu ,\nu \right) }(1-2\exp (-2\alpha r))=\frac{\left( \mu
+1\right) _{n}}{n!}%
\begin{array}{c}
_{2}F_{1}%
\end{array}%
\left( -n,1+\mu +\nu +n;\mu +1;\exp (-2\alpha r)\right) ,
\end{equation}%
where $z\in \left[ 0,1\right] $ which lie within or on the boundary of the
interval $\left[ -1,1\right] .$

Fourth, if we choose $V_{2}\rightarrow iV_{2},$ the potential becomes the $%
PT $-symmetric Rosen-Morse potential, where $P$ denotes parity operator and $%
T$ denotes time reversal. \ For a potential $V(r),$ making the
transformation of $r-r$ (or $r\rightarrow \xi -r$) and $i\rightarrow -i,$ if
we have the relation $V(-r)=V^{\ast }(r)),$ the potential $V(r)$ is said to
be $PT$-symmetric [43]. In this case we obtain for spin-symmetric Dirac
equation%
\begin{equation*}
\left( Mc^{2}+E_{n\kappa }\right) \left( Mc^{2}-E_{n\kappa }+iV_{2}\right) =-%
\frac{\omega D_{0}}{r_{e}^{2}}\hbar ^{2}c^{2}
\end{equation*}%
\begin{equation}
+\alpha ^{2}\hbar ^{2}c^{2}\left[ \frac{-\frac{iV_{2}}{2\alpha ^{2}\hbar
^{2}c^{2}}\left( Mc^{2}+E_{n\kappa }-C_{s}\right) +\frac{\omega \left(
D_{1}+D_{2}\right) }{4\alpha ^{2}r_{e}^{2}}}{\left( n+\widetilde{\delta }%
+1\right) }-\left( n+\widetilde{\delta }+1\right) \right] ^{2}.
\end{equation}%
In the non-relativistic limit, it turns to become%
\begin{equation}
E_{nl}=iV_{2}+\frac{\omega \hbar ^{2}D_{0}}{2\mu r_{e}^{2}}-\frac{\hbar ^{2}%
}{2\mu }\alpha ^{2}\left[ \frac{\frac{\mu }{\alpha ^{2}\hbar ^{2}}\left(
2V_{1}+iV_{2}\right) -\frac{\omega D_{1}}{4\alpha ^{2}r_{e}^{2}}+\left(
n+1\right) ^{2}+\left( 2n+1\right) \widetilde{\delta }_{0}}{n+\widetilde{%
\delta }_{0}+1}\right] ^{2},
\end{equation}%
where real $V_{1}>0,$ which is identical to the results of Ref. [25]. If one
sets $l=0$ in the above equation, the result is identical with that of Refs.
[44,45].

\section{Conclusions}

We have obtained analytically the energy spectra and corresponding wave
functions of the Dirac equation for the Rosen-Morse potential under the
conditions of the spin symmetry and pseudospin symmetry in the context of
the Nikiforov-Uvarov method. For any spin-orbit coupling centrifugal term $%
\kappa ,$ we have found the explicit expressions for energy eigenvalues and
associated wave functions in closed form. The most stringent interesting
result is that the present spin and pseudospin symmetry cases can be easily
reduced to the KG solution once $S(r)=V(r)$ and $S(r)=-V(r)$ (\textit{i.e}., 
$C_{s}=C_{ps}=0$) [55]. The resulting solutions of the wave functions are
being expressed in terms of the generalized Jacobi polynomials. Obviously,
the relativistic solution can be reduced to it's non-relativistic limit by
the choice of appropriate mapping transformations. Also, in case when
spin-orbit quantum number $\kappa =0,$ the problem reduces to the $s$-wave
solution. The $s$-wave Rosen-Morse, the Eckart-type potential, the
PT-symmetric Rosen-Morse potential.and the non-relativistic cases are
briefly studied.

\acknowledgments The partial support provided by the Scientific and
Technological Research Council of Turkey (T\"{U}B\.{I}TAK) is highly
appreciated. The author thanks the anonymous kind referees and editors for
the very constructive comments and suggestions.

\newpage \appendix

\section{Parametric Generalization of the NU Method}

Our systematical derivation holds for any potential form.

(i) The relevant coefficients $c_{i}$ ($i=5,6,\cdots ,16$) are given as
follows:%
\begin{equation}
c_{5}=\frac{1}{2}\left( c_{3}-c_{1}\right) ,\text{ }c_{6}=\frac{1}{2}\left(
c_{2}-2c_{4}\right) ,\text{ }c_{7}=c_{6}^{2}+\xi _{1},
\end{equation}%
\begin{equation}
\text{ }c_{8}=2c_{5}c_{6}-\xi _{2},\text{ }c_{9}=c_{5}^{2}+\xi _{3},\text{ }%
c_{10}=c_{4}\left( c_{3}c_{8}+c_{4}c_{9}\right) +c_{3}^{2}c_{7},
\end{equation}%
\begin{equation}
c_{11}=\frac{2}{c_{3}}\sqrt{c_{9}},\text{ }c_{12}=\frac{2}{c_{3}c_{4}}\sqrt{%
c_{10}},
\end{equation}%
\begin{equation}
c_{13}=\frac{1}{c_{3}}\left( c_{5}+\sqrt{c_{9}}\right) ,\text{ }c_{14}=\frac{%
1}{c_{3}c_{4}}\left( \sqrt{c_{10}}-c_{4}c_{5}-c_{3}c_{6}\right) ,
\end{equation}%
\begin{equation}
c_{15}=\frac{2}{c_{3}}\sqrt{c_{10}},\text{ }c_{16}=\frac{1}{c_{3}}\left( 
\sqrt{c_{10}}-c_{4}c_{5}-c_{3}c_{6}\right) .
\end{equation}%
(ii) The analytic results for the key polynomials: 
\begin{equation}
\pi (r)=c_{5}+\sqrt{c_{9}}-\frac{1}{c_{3}}\left( c_{4}\sqrt{c_{9}}+\sqrt{%
c_{10}}-c_{3}c_{6}\right) r,
\end{equation}%
\begin{equation}
k=-\frac{1}{c_{3}^{2}}\left( c_{3}c_{8}+2c_{4}c_{9}+2\sqrt{c_{9}c_{10}}%
\right) ,
\end{equation}%
\begin{equation}
\tau (r)=c_{3}+2\sqrt{c_{9}}-\frac{2}{c_{3}}\left( c_{3}c_{4}+c_{4}\sqrt{%
c_{9}}+\sqrt{c_{10}}\right) r,
\end{equation}%
\begin{equation}
\tau ^{\prime }(r)=-\frac{2}{c_{3}}\left( c_{3}c_{4}+c_{4}\sqrt{c_{9}}+\sqrt{%
c_{10}}\right) <0.
\end{equation}%
(iii) The energy equation:%
\begin{equation}
c_{2}n-\left( 2n+1\right) c_{6}+\frac{1}{c_{3}}\left( 2n+1\right) \left( 
\sqrt{c_{10}}+c_{4}\sqrt{c_{9}}\right) +n\left( n-1\right) c_{4}+\frac{1}{%
c_{3}^{2}}\left( c_{3}c_{8}+2c_{4}c_{9}+2\sqrt{c_{9}c_{10}}\right) =0.
\end{equation}%
(iv) The wave functions:%
\begin{equation}
\rho (r)=r^{c_{11}}(c_{3}-c_{4}r)^{c_{12}},
\end{equation}%
\begin{equation}
\phi (r)=r^{c_{13}}(c_{3}-c_{4}r)^{c_{14}},\text{ }c_{13}>0,\text{ }c_{14}>0,
\end{equation}%
\begin{equation}
y_{n\kappa }(r)=P_{n}^{\left( c_{11},c_{12}\right) }(c_{3}-2c_{4}r),\text{ }%
c_{11}>-1,\text{ }c_{12}>-1,\text{ }r\in \left[
(c_{3}-1)/2c_{4},(1+c_{3})/2c_{4}\right] ,
\end{equation}%
\begin{equation}
\psi _{n\kappa }(r)=\phi (r)y_{n\kappa }(r)=\mathcal{N}%
_{n}r^{c_{13}}(c_{3}-c_{4}r)^{c_{14}}P_{n}^{\left( c_{11},c_{12}\right)
}(c_{3}-2c_{4}r),
\end{equation}%
where $P_{n}^{\left( a,b\right) }(c_{3}-2c_{4}r)$ are the Jacobi polynomials
and $\mathcal{N}_{n}$ is a normalizing factor.

When $c_{4}=0,$ the Jacobi polynomial turn to be the generalized Laguerre
polynomial and the constants relevant to this polynomial change are%
\begin{equation}
\lim_{c_{4}\rightarrow
0}P_{n}^{(c_{11},c_{12})}(c_{3}-2c_{4}r)=L_{n}^{c_{11}}(c_{15}r),
\end{equation}%
\begin{equation}
\lim_{c_{4}\rightarrow 0}(c_{3}-c_{4}r)^{c_{14}}=\exp (-c_{16}r),
\end{equation}%
\begin{equation}
\psi _{n\kappa }(r)=\mathcal{N}_{n}\exp (-c_{16}r)L_{n}^{c_{11}}(c_{15}r),
\end{equation}%
where $L_{n}^{c_{11}}(c_{15}r)$ are the generalized Laguerre polynomials and 
$\mathcal{N}_{n}$ is a normalizing constant.

$\label{appendix}$

\section{Normalization of the radial wave function}

In order to find the normalization factor $\mathcal{N}_{n\kappa }$, we start
by writting the normalization condition:%
\begin{equation}
\frac{\mathcal{N}_{n\kappa }^{2}}{2\alpha }\int_{0}^{1}z^{2\varepsilon
_{n\kappa }-1}(1-z)^{2\delta +2}\left[ P_{n}^{(2\varepsilon _{n\kappa
},2\delta +1)}(1-2z)\right] ^{2}dz=1.
\end{equation}%
Unfortunately, there is no formula available to calculate this key
integration. Neveretheless, we can find the explicit normalization constant $%
\mathcal{N}_{nl}.$ For this purpose, it is not difficult to obtain the
results of the above integral by using the following formulas [61-64]%
\begin{equation}
\dint\limits_{0}^{1}\left( 1-s\right) ^{\mu -1}s^{\nu -1}%
\begin{array}{c}
_{2}F_{1}%
\end{array}%
\left( \alpha ,\beta ;\gamma ;as\right) ds=\frac{\Gamma (\mu )\Gamma (\nu )}{%
\Gamma (\mu +\nu )}%
\begin{array}{c}
_{3}F_{2}%
\end{array}%
\left( \nu ,\alpha ,\beta ;\mu +\nu ;\gamma ;a\right) ,
\end{equation}%
and $%
\begin{array}{c}
_{2}F_{1}%
\end{array}%
\left( a,b;c;z\right) =\frac{\Gamma (c)}{\Gamma (a)\Gamma (b)}%
\dsum\limits_{p=0}^{\infty }\frac{\Gamma (a+p)\Gamma (b+p)}{\Gamma (c+p)}%
\frac{z^{p}}{p!}.$ Hence, the normalization constants for the upper-spinor
component are%
\begin{equation}
\mathcal{N}_{n\kappa }=\left[ \frac{\Gamma (2\delta +3)\Gamma (2\varepsilon
_{n\kappa }+1)}{2\alpha \Gamma (n)}\dsum\limits_{m=0}^{\infty }\frac{%
(-1)^{m}\left( n+2(1+\varepsilon _{n\kappa }+\delta )\right) _{m}\Gamma (n+m)%
}{m!\left( m+2\varepsilon _{n\kappa }\right) !\Gamma \left( m+2\left(
\varepsilon _{n\kappa }+\delta +\frac{3}{2}\right) \right) }f_{n\kappa }%
\right] ^{-1/2}\text{ ,}
\end{equation}%
with 
\begin{equation}
f_{n\kappa }=%
\begin{array}{c}
_{3}F_{2}%
\end{array}%
\left( 2\varepsilon _{n\kappa }+m,-n,n+2(1+\varepsilon _{n\kappa }+\delta
);m+2\left( \varepsilon _{n\kappa }+\delta +\frac{3}{2}\right)
;1+2\varepsilon _{n\kappa };1\right) ,
\end{equation}%
where $\left( x\right) _{m}=\Gamma (x+m)/\Gamma (x).$ Also, the
normalization constants for the lower-spinor component are%
\begin{equation}
\widetilde{\mathcal{N}}_{n\kappa }=\left[ \frac{\Gamma (2\delta
_{1}+3)\Gamma (2\widetilde{\varepsilon }_{n\kappa }+1)}{2\alpha \Gamma (n)}%
\dsum\limits_{m=0}^{\infty }\frac{(-1)^{m}\left( n+2(1+\widetilde{%
\varepsilon }_{n\kappa }+\delta _{1})\right) _{m}\Gamma (n+m)}{m!\left( m+2%
\widetilde{\varepsilon }_{n\kappa }\right) !\Gamma \left( m+2\left( 
\widetilde{\varepsilon }_{n\kappa }+\delta _{1}+\frac{3}{2}\right) \right) }%
g_{n\kappa }\right] ^{-1/2}\text{ ,}
\end{equation}%
with 
\begin{equation}
g_{n\kappa }=%
\begin{array}{c}
_{3}F_{2}%
\end{array}%
\left( 2\widetilde{\varepsilon }_{n\kappa }+m,-n,n+2(1+\widetilde{%
\varepsilon }_{n\kappa }+\delta _{1});m+2\left( \widetilde{\varepsilon }%
_{n\kappa }+\delta _{1}+\frac{3}{2}\right) ;1+2\widetilde{\varepsilon }%
_{n\kappa };1\right) .
\end{equation}%
\newpage\ 

\ {\normalsize 
}

\bigskip

\bigskip

\baselineskip= 2\baselineskip
\bigskip \newpage

\bigskip

\bigskip {\normalsize 
}

\baselineskip= 2\baselineskip

\bigskip 
\begin{table}[tbp]
\caption{The specific values for the parametric constants necessary for
calculating the energy eigenvalues and eigenfunctions of the spin symmetry
Dirac wave equation.}%
\begin{tabular}{llll}
\tableline Constant & Analytic value & Constant & Analytic value \\ 
\tableline$c_{1}$ & $1$ & $c_{2}$ & $1$ \\ 
$c_{3}$ & $1$ & c$_{4}$ & $1$ \\ 
$c_{5}$ & $0$ & $c_{6}$ & $-\frac{1}{2}$ \\ 
$c_{7}$ & $\frac{1}{4}+\beta _{1}$ & $c_{8}$ & $-\beta _{2}$ \\ 
$c_{9}$ & $\varepsilon _{n\kappa }^{2}$ & $c_{10}$ & $\left( \delta +\frac{1%
}{2}\right) ^{2}$ \\ 
$c_{11}$ & $2\varepsilon _{n\kappa }$ & $c_{12}=c_{15}$ & 2$\delta +1$ \\ 
$c_{13}$ & $\varepsilon _{n\kappa }$ & $c_{14}=c_{16}$ & $\delta +1$ \\ 
$\xi _{1}$ & $\beta _{1}$ & $\xi _{2}$ & $\beta _{2}$ \\ 
$\xi _{3}$ & $\varepsilon _{n\kappa }^{2}$ &  &  \\ 
\tableline &  &  & 
\end{tabular}%
\end{table}
\ 

\end{document}